# Phonon vortices at heavy impurities in two-dimensional materials

De-Liang Bao[1], Mingquan Xu[2], Ao-Wen Li[2], Gang Su[2,3], Wu Zhou[2], and Sokrates T. Pantelides[1,4,*]

[1]*Department of Physics and Astronomy, Vanderbilt University, Nashville, Tennessee 39235, U.S.A.*
[2]*School of Physical Sciences and CAS Key Laboratory of Vacuum Physics, University of the Chinese Academy of Sciences, Beijing 100049, P. R. China*
[3]*Kavli Institute for Theoretical Sciences, University of the Chinese Academy of Sciences, Beijing, P. R. China*
[4]*Department of Electrical and Computer Engineering, Vanderbilt University, Nashville, Tennessee 37235, U.S.A.*

The advent of monochromated electron energy-loss spectroscopy has enabled atomic-resolution vibrational spectroscopy, which triggered interest in spatially localized or quasi-localized vibrational modes in materials. Here we report the discovery of phonon vortices at heavy impurities in two-dimensional materials. We use density-functional-theory calculations for two configurations of Si impurities in graphene, Si-$C_3$ and Si-$C_4$, to examine atom-projected phonon densities of states and display the atomic-displacement patterns for select modes that are dominated by impurity displacements. The vortices are driven by large displacements of the impurities, and reflect local symmetries. Similar vortices are found at phosphorus impurities in hexagonal boron nitride, suggesting that they may be a feature of heavy impurities in crystalline materials. Phonon vortices at defects are expected to play a role in thermal conductivity and other properties.

*pantelides@vanderbilt.edu

Defects, namely inhomogeneities in the atomic arrangement of materials, are ubiquitous, making their engineering essential to tune macroscopic functional properties. Extensive fundamental research has been conducted on defects, focusing on topics such as electronic energy levels in semiconductor band gaps[1], defect formation and migration energies[2,3], optical and magnetic properties[4], catalytic effects[5], and defect-mediated reliability issues in electronic, optoelectronic, and electrical devices[6-8]. Nevertheless, the vibrational properties of defects have received limited attention even though they are equally fundamental and play a role in thermal conduction[9], phase transitions[10], structural stability[11], infrared absorption[12] and dielectric response[13], nanofluid transport[14], and superconductivity through electron-phonon coupling[15]. Conventional treatments of phonons in defective materials typically focus on phonon scattering from defects[16], which affects parameters such as phonon lifetimes, mean free paths, and phonon-phonon coupling. Unusual phenomena have also been found to occur at defects, such as the stabilization of carbon-hydrogen bonds on graphene by an in-plane sonic wave[17] and the use of acoustic-phonon waves to probe the location of point defects[18].

Conventionally, experimental investigations of phonons in materials have been based on optical and neutron-scattering techniques with limited spatial resolution. Recent advances in experimental techniques, however, particularly the development of monochromated electron energy-loss spectroscopy (EELS) in scanning transmission electron microscopes (STEM), have revolutionized this field[19]. These breakthroughs paved the way for phonon spectroscopy with exceptional atomic-scale spatial resolution and meV-scale energy resolution[20]. Atomic-scale phonon spectroscopy has been carried out at point defects[21], interfaces[22], stacking faults[23], grain boundaries[24], and even with chemical-bond sensitivity[25]. The spatially resolved spectra feature peaks at energies that are distinct from those of the pure material, arising from different atomic masses and local atomic arrangements[21,25]. Theoretical calculations based on density-functional theory (DFT) have been an equal partner in the enterprise by offering predictions of such energies and going beyond the reach of experiments to describe the eigen-displacements and the extent of localization, which depends on whether the new energies lie within the continuum or within energy gaps in perfect-crystal phonon dispersions[21-25].

In this paper, we employ DFT calculations and unveil a particularly notable feature of the vibrational properties of point defects in two-dimensional materials, namely the existence of vibrational modes with atomic displacements that form circular patterns around a heavy impurity in the shape of a vortex. We call such circular vibrational patterns "phonon vortices". The energies (frequencies) of these modes do not necessarily lie in an energy gap in the perfect-crystal phonon dispersion, even if such a gap exists, whereby the corresponding vortices are connected to perfect-crystal phonon waves (such defect-induced quasi-localized states within the band continua are known to be induced by defects in the electron energy bands and are referred to as resonances[26]). Thus, defect-induced phonon vortices within the continua of perfect-crystal phonon dispersion are analogs of vortices that form when flowing water runs into an obstacle, but the "obstacle" here, namely a heavy impurity atom, is actually vibrating with a relatively large displacement. More specifically, we report results for impurities in graphene and two-dimensional (2D) hexagonal boron nitride (h-BN). In graphene, we focus on Si impurities as prototype systems because Si is 2.3 times as heavy as C and is known to form two point defects with distinct bonding configurations, namely Si-C$_3$ (a Si atom replacing a single C atom) and Si-$C_4$ (a Si atom replacing two nearest-neighbor C atoms), shown in schematics in Figure 1a. We follow up on the findings in Ref. 25, which tracked the atom-by-atom vibrational EELS (vEELS) response at these two Si bonding configurations. The corresponding DFT calculations of atom-projected phonon density of states (PhDOS) successfully reproduce the vibrational features in measured vEELS and provide detailed analysis of the origin and shifts of the spectral peaks. The STEM/EELS energy resolution is not currently sufficient to map the vibrational patterns of individual frequencies, allowing only averages over a small range of frequencies, but future mapping of vortices remains a possibility. Calculations of phonon vortices in h-BN are also reported, but calculations in 2D materials with multiple atomic planes or bulk crystals are still prohibitively expensive. Phonon vortices at impurities are likely to impact phonon-impurity scattering and thus thermal properties such as a thermal conductance. For the same reason, even slight intermixing at interfaces may impact thermal conductivity.

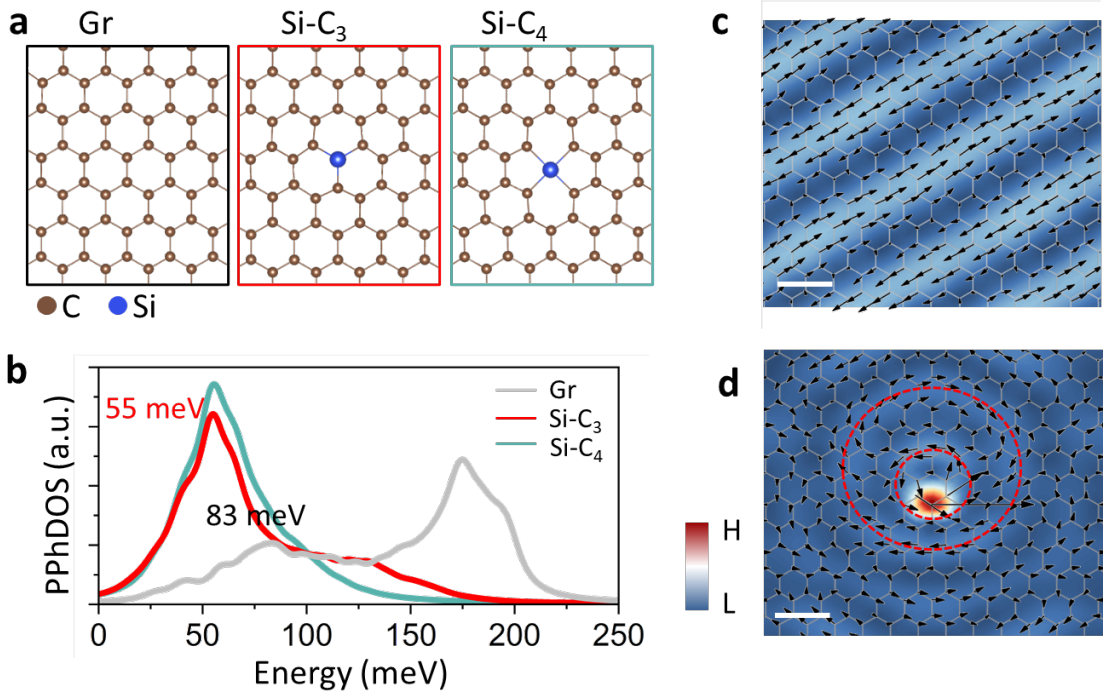

**Figure 1. Phonons in intrinsic graphene and at defects. a**. Atomic schematics of intrinsic graphene (left), substitutional Si in graphene, Si-$C_3$ (middle), and Si-$C_4$, where the Si sits in a divacancy and therefore has four nearest C atoms (right). **b**. In-plane projected phonon density of states (PPhDOS) on one C atom in graphene (gray curve), on the Si impurity in Si-$C_3$ (red curve), and on the Si impurity in Si-$C_4$ (cyan curve). **c**. Eigenvector map of a typical phonon mode of intrinsic graphene at an energy of 39 meV. **d**. Eigenvector map of the vortex phonon mode of Si-$C_3$ at the energy of 55 meV (peak energy in panel **b**). In the maps shown in **c** and **d**, the arrows represent the in-plane components of atomic eigen-displacements, and the colors correspond to the amplitude of atomic displacements. Scale bars: 0.5 nm.

Graphene, a 2D honeycomb carbon network (Fig. 1a, left), has recently served as an ideal platform for studying phonons using monochromated vibrational EELS[21,25]. With a relatively large broadening of the eigenstates, the phonon density of states (PhDOS) of graphene features two primary peaks. The lower-energy peak at ~83 meV derives mostly from acoustic modes, while the higher-energy peak at ~175 meV derives from optical modes[25]. As expected, the projected PhDOS (PPhDOS) on a single C atom (gray in Fig. 1b) looks essentially identical as the total PhDOS[21].

We examine phonons in graphene in the presence of the Si-$C_3$ and Si-$C_4$ defects that are depicted schematically in Fig 1a. In both cases, the PPhDOS on Si (red and green curves in Fig 1b) shows that a low-energy peak at ~55 meV is induced while the optical peak effectively vanishes. These PPhDOSs on Si in Si-$C_3$ and Si-$C_4$ have been validated by atomically-resolved vEELS experiments[21,25]. We attribute the distinct PPhDOS behavior of Si to the heavier mass of Si. In the acoustic region, the C atoms move collectively along the direction of the eigenvectors. The heavier Si vibrates intensely at a relatively low energy, according to the harmonic oscillator formula $E = \hbar\omega = \hbar\sqrt{k/m}$, where $\omega$ is the frequency of vibration, $k$ is an effective spring constant, and $m$ is the Si-atom mass. The disappearance of the optical peak at 175 meV in the PPhDOS of Si is also caused by the Si-atom heavy mass. Si atoms simply cannot vibrate at such high frequencies. Recall that for perfect-crystal SiC, the phonon energies are no higher than 125 meV[27] while in crystalline Si they only range to 65 meV[28].

Given the notable distinctions in the PPhDOS on Si and on C in pristine graphene, the investigation of atomic eigen-displacements around defects becomes particularly intriguing. Figure 1c depicts the eigen-displacement map of an individual phonon mode of pristine graphene at 39 meV, displaying a linear pattern with a periodicity corresponding to the wavelength. In contrast, the eigen-displacements map of Si-$C_3$ around the first primary peak at ~55 meV shows a remarkably different pattern (Fig. 1d). First, the color map, representing the atomic displacement amplitude, indicates a very high intensity at the Si impurity, with observable intensity also present at the neighboring C atoms, suggesting that it is an impurity-dominated resonant phonon mode[29]. Second, the color map shows that the local atomic displacements form circular features around the Si atom and dispersive strips radiating away from the Si. We have inserted two red dashed circles to highlight the dominant circular features. The inner circle passes very near the Si impurity, while the outer circle surrounds the inner one. The arrows depicted on the red circles represent a vortex behavior. The motion of the atoms is back and forth, in concert, in the direction of the little arrows.

To explore the formation mechanism of the phonon vortex, we reproduced the vortex in Fig. 1d using a larger computational supercell (Fig. 2a) and sandwiched it between two side panels showcasing the atomic displacements of the pristine-graphene mode closest to the energy at which the vortex forms. The intentionally placed gaps between the middle and side panels indicate that the connection to the pristine-graphene mode is likely to occur occurs further out (see discussion later in the paper). In the middle panel, beyond the two inner vortices (red dashed circles in Fig. 2a), there are bending curves (yellow dashed curves in Fig. 2a). Upon closer examination, it becomes evident that the yellow dashed curves in the middle panel align remarkably well with the solid yellow lines in the left and right panels. This alignment strongly suggests that the yellow dashed curves correspond to distorted phonon waves in intrinsic graphene. Recalling that the PPhDOS on Si of the Si-$C_3$ defect only shows a primary peak in the acoustic-phonon regime of graphene, the formation of a phonon vortex in Si-$C_3$ arises from the coupling between Si-atom vibrations and the pristine-graphene Bloch acoustic-phonon at the same energy. To illustrate the mechanism, we draw an analogy between the flow of water around an obstacle (right inset in Fig. 2b) and the formation of the phonon vortex. When water waves, analogous to phonon waves, propagate around an obstacle (representing the defect), a vortex forms around the obstacle (defect). Accordingly, the vortices depend on the obstacle's shape and size (analogous to the defect's local symmetry and mass, respectively), as well as the properties of the flowing liquid (representing the host material).

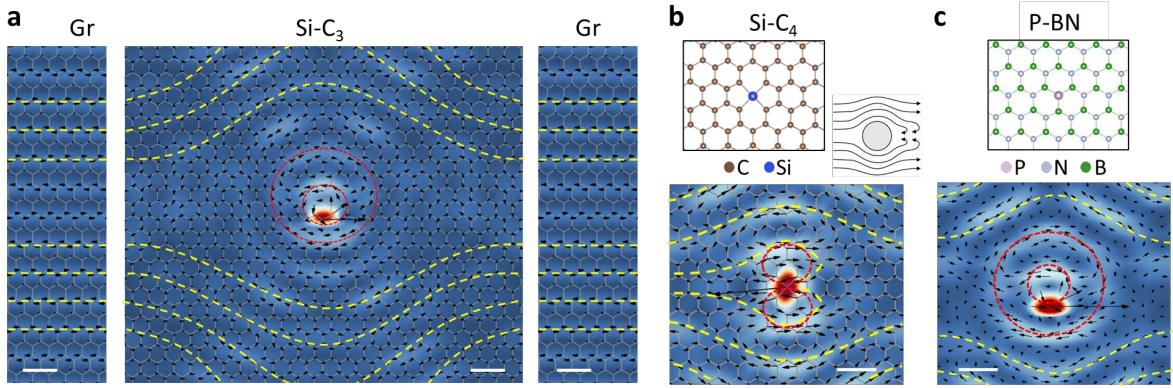

**Figure 2. Phonon vortices at defects**. **a**. A larger view of the Si-$C_3$ vortex of Fig. 1d, obtained with a four-times-larger supercell (see Methods), sandwiched between side panels depicting the eigen-displacements of a pristine-graphene mode at ~57 meV, which is closest to the energy of the PPhDOS on Si in the larger supercell. The gaps between the panels indicate that the overall connection is likely to happen farther away from the vortex. **b**. A phonon-vortex map of Si-$C_4$ at 56 meV. Top inset: Atomic configuration of Si-$C_4$ and illustration of water flowing around an obstacle. **c**. A phonon-vortex map of a phonon mode at ~46 meV at a P atom substituting a N atom in h-BN. The dashed red circles and dashed yellow curves in all plots mark the traces of atoms with relatively big displacement amplitudes in corresponding phonon modes. Scale bars: 0.5 nm.

The morphology of phonon vortices relates to the local symmetry of defects. In Si-$C_4$, the Si impurity is chemically bonded to four surrounding C atoms. We again observe phonon vortices in Si-$C_4$ at the phonon mode with the frequency around the first primary peak of the PPhDOS on Si. The Si-$C_4$ phonon vortex (Fig. 2b) features a butterfly-like morphology, which arises from the local mirror symmetry inherent in the Si-$C_4$ defect. In contrast, the Si-$C_3$ defect retains the local trigonal symmetry of C atoms and the vortex shown in Fig. 2a represents one of three possible orientations that the vortex can have. Clearly, the symmetry of the defect significantly influences the shape and behavior of the phonon vortices.

Phonon vortices at point defects are present in other host materials. Monolayer h-BN is another honeycomb 2D material that consists of two different sublattices occupied by B and N atoms, respectively. It is worth noting that B and N atoms in h-BN possess different Born effective charges, leading to the presence of infrared-active phonon modes that can generate electronic polarizations through eigen-displacements. The presence of a phonon vortex in h-BN would be more interesting with respect to dielectric properties. When one N atom is replaced by a heavier P atom, the formation of a phonon vortex similar to that of Si-$C_3$ is observed (Fig. 2d), which suggests a common underlying mechanism for the generation of vortices in graphene and h-BN. The frequency of phonon vortices in P-BN is within the acoustic-phonon regime of h-BN, *i.e.*, it is also a resonant phonon mode as is the case for Si defects in graphene.

It is clear from the eigen-displacement maps of the vortices in Figs. 1 and 2 that the vortices are larger than the computational supercells, which raises questions about the effect of neighboring vortices on the vortex shape. Si-$C_3$ vortices in adjacent 12×12 computational supercells are shown in Fig. 3a. That is the supercell used for the PPhDOS curves and the vortex map in Figs. 1b and 1d, respectively. Though computationally taxing, we performed calculations in which we quadrupled the supercell size for Si-$C_3$ (the maps of the Si-$C_3$ vortex in Fig. 1d and Fig. 2a are from 12×12 and 24×24 supercells, respectively, but are placed in a square periodic unit). The results shown in Fig. 3b reveal that the morphology and size of the vortex cores do not change noticeably. The dashed rhombi in Fig. 3b act as reference markers, guiding the readers' eyes to recognize and compare the similarities in the vortex patterns observed within the respective supercells. In computational parlance, *the vortex-core morphology is converged*. The fact that the vortex appears in the graphene mode with energy at the primary peak in the PPhDOS on Si (Fig. 1b) corroborates this conclusion.

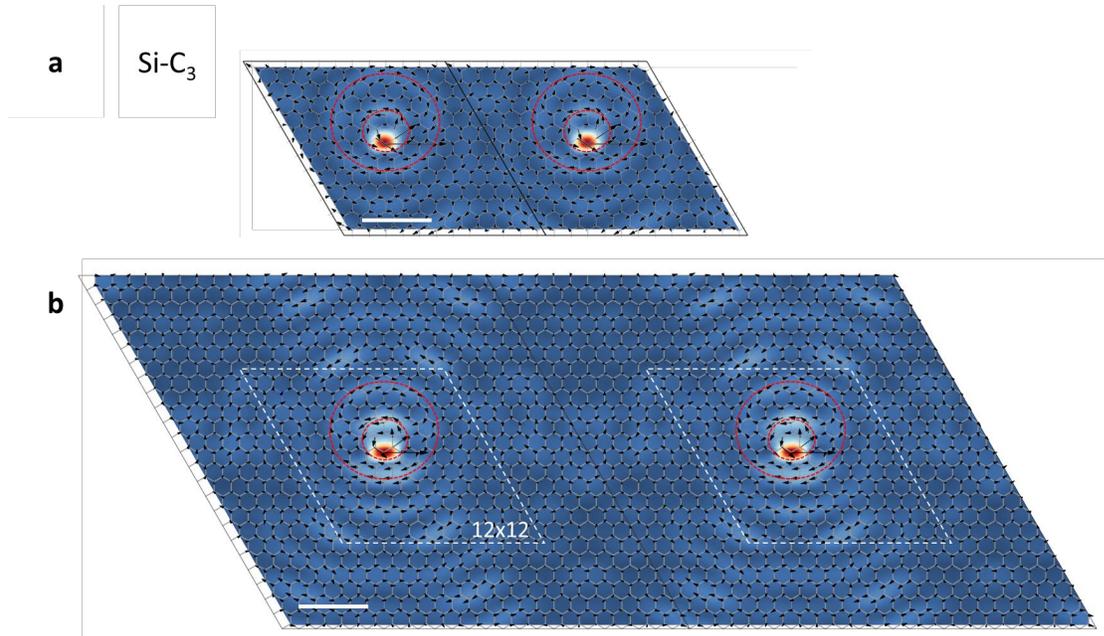

**Figure 3. Phonon vortices in computational supercells of different sizes. a**. Eigen-displacement maps of Si-$C_3$ phonon vortices in two adjacent 12×12 computational supercells. Red circles again illustrate the phonon vortices. The energy of the mode is 55 meV. **b**. Eigen-displacement maps of Si-$C_3$ phonon vortices in two adjacent 24×24 computational supercells. The frequency of the plotted mode is 57 meV, matching the energy of the peak in the PPhDOS obtaind with 24×24 supercells. The two white dashed rhombi guide the eyes to identify the similarities with the two rhombi in Fig. 3a. Scalebars: 1 nm.

On the other hand, Fig. 3b makes clear that the red circular patterns around the Si impurity continue outside the dashed romboids, *i.e.*, the overall size of the vortex is much larger than what we see in the central panel of Fig. 1d. It is also clear that the distance between vortices is not large enough to allow the presence of pristine-graphene phonon waves, as we illustrated schematically in the side panels of Fig. 2a. Much larger supercells are needed for such features to materialize. Such calculations are not currently practical.

The vortices we described in this paper are unique to point defects, and are fundamentally different from chiral phonon waves[30,31]. In a phonon vortex at a defect, all atoms execute small back-and-forth motions, in concert, in the directions indicated by the black arrows in our vortex maps, as in the case of conventional acoustic and optical phonons in crystals, but the arrows in a vortex are arranged in a roughly circular pattern around the defect. The atomic displacements of chiral phonons are not back and forth along an axis. Another difference is in the angular momentum of chiral phonons and vortices. All atoms participating in vortices vibrate back and forth along an axis, so that the net angular momentum about the central axis is zero (the contribution to the angular momentum of each atom is *r*×*p* half of the time and *r*×(-*p*) the other half of the time). In contrast, atoms in chiral phonons circle their equilibrium sites and have non-zero angular momentum, while the net angular momentum of the chiral mode depends on the relative phase of rotation of atoms in different sublattices[30,31].

In summary, we performed extensive calculations of phonons at various point defects anchored by a heavy impurity in different 2D monolayer materials and found phonon vortices. The phonon vortices are characterized by back-and-forth displacements in concert in circular patterns. The formation of phonon vortices is attributed to the coupling between the vibrations of the heavy impurity atom and the propagating acoustic phonon waves of the host monolayer. We demonstrated that the behavior of phonon vortices depends on the local symmetry and mass of the defect and that phonon vortices at point defects exist in different host materials, namely graphene and h-BN. The question whether phonon vortices at point defects exist in 2D materials that consist of more than one plane of atoms and in three-dimensional materials remains an open question at this stage as accurate calculations are not practical. When vortices do exist, however, they are likely to impact thermal properties, especially thermal conductivity.

**Methods**

The DFT calculations were performed using the Vienna ab initio Simulation Package (VASP)[32,33] with the projector-augmented-wave method[34]. The Perdew-Burke-Ernzerhof[35] version of the generalized gradient approximation (GGA)[36] was used in all simulations. The plane-wave basis energy cut off is 600 eV. The calculational cells for Si- and N-doped graphene are 12×12 supercells, containing 288 atomic sites. Some calculations were done 24×24 supercells containing 1152 atomic sites. Calculations on substitutional P in hBN were performed using 12×12 supercells, with one P atom substituting a N atom. A vacuum region between graphene (or BN) sheets was set to 20 Å along the $z$ direction to minimize periodic interactions. The **k** points were set as Γ-centered 3×3×1 for structural optimization and 1×1×1 for phonon calculations due to the expensive computational cost of the latter. For structural relaxation, the structures were relaxed until the atomic forces were less than 0.001 eV/Å. Phonon calculations were performed using VASP combined with Phonopy using the density-functional-perturbation-theory (DFPT) method. For supercells, only Γ-point phonons were considered, while for pristine graphene, necessary $q$ points over the Brillouin zone were considered. The projected phonon densities of states are projected in the graphene (or BN) plane on corresponding atoms. A full width at half maximum (FWHM) of 8 meV was used to plot the projected phonon density of states. The arrows presenting eigenvectors are the in-plane components of corresponding phonon modes. The Si-C$_4$ structure features four Si-C bonds with a slightly buckled configuration (the four C neighbors of Si rise by 0.2 Å above the graphene basal plane), whereas the Si-C$_3$ features three Si-C bonds that are buckled (Si rises 1.8 Å above the graphene basal plane).

**Conflicts of interest**

There are no conflicts to declare.

**Acknowledgements**

Research at Vanderbilt University (D.-L.B., S.T.P.) was supported by the U.S. Department of Energy, Office of Science, Basic Energy Sciences, Materials Science and Engineering Division Grant No. DE-FG02-09ER46554 and by the McMinn Endowment. Calculations were performed at the National Energy Research Scientific Computing Center (NERSC), a U.S. Department of Energy Office of Science User Facility located at Lawrence Berkeley National Laboratory, operated under Contract No. DE-AC02-05CH11231.

Research at UCAS was supported by the Beijing Outstanding Young Scientist Program with grant No. BJJWZYJH01201914430039 (W.Z.).